\documentclass{emulateapj}

\slugcomment{Accepted by ApJL on 2009 May 21}

\shorttitle{Candidate Binary Black-Hole System}
\shortauthors{Wrobel \& Laor}

\begin{document}

\title{Discovery of Radio Emission from the Quasar 
       SDSS J1536+0441, a Candidate
       Binary Black-Hole System}

\author{J. M. Wrobel\altaffilmark{1} and A. Laor\altaffilmark{2}}

\altaffiltext{1}{National Radio Astronomy Observatory, P.O. Box O,
Socorro, NM 87801; jwrobel@nrao.edu}
\altaffiltext{2}{Physics Department, Technion, Haifa 32000, Israel;
laor@physics.technion.ac.il}

\begin{abstract}
The radio-quiet quasar SDSS J1536+0441 shows two broad-line emission
systems that Boroson \& Lauer interpret as a candidate binary
black-hole system with a separation of 0.1 pc (0.02 mas).  From new
VLA imaging at 8.5 GHz, two faint sources, separated by 0.97\arcsec\,
(5.1 kpc), have been discovered within the quasar's optical
localization region.  Each radio source is unresolved, with a diameter
of less than 0.37\arcsec\, (1.9 kpc).  A double radio structure is
seen in some other radio-quiet quasars, and the double may be
energized here by the candidate 0.1-pc binary black-hole system.
Alternatively, the radio emission may arise from a binary system of
quasars with a projected separation of 5.1 kpc, and the two quasars
may produce the two observed broad-line emission systems.  Binary
active galactic nuclei with a kpc scale separation are known from
radio and X-ray observations, and a few such system are expected in
the Boroson \& Lauer sample based on the observed clustering of
quasars down to the 10 kpc scale.  Future observations designed to
distinguish between the 0.1 pc and 5 kpc scales for the binary system
are suggested.
\end{abstract}

\keywords{quasars: individual (SDSS J1536+0441) ---
          radio continuum: general ---
          X-rays: general}

\section{Motivation}\label{motivation}

The quasar \object{SDSS J153636.22+044127.0}, at a redshift 0.388,
shows two broad-line emission systems separated in velocity by 3500
km~s$^{-1}$ \citep{bor09}.  This unique quasar is interpreted as a
binary black-hole system with a separation of 0.1 pc and masses of
$10^{7.3}~M_\sun$ and $10^{8.9}~M_\sun$.  The sub-parsec scale is
significant, as it suggests that this close binary system shares a
common narrow-line emission region and has also solved its so-called
final parsec problem.  For the assumed flat cosmology\footnote{$H_0 =
71$~km~s$^{-1}$~Mpc$^{-1}$ and $\Omega_m = 0.27$, implying a
luminosity distance of 2.1 Gpc, an angular size distance of 1.1 Gpc
and a scale of 5.2 kpc per arc sec.}, the 0.1-pc binary subtends 0.02
mas.  An alternative interpretation for SDSS J153636.22+044127.0 (SDSS
J1536+0441 hereafter) is that it represents two unrelated quasars
viewed, by chance, along similar sight lines.  \citet{bor09} use the
optical localization region - a circle of radius 1\arcsec\, - to rule
out this alternative interpretation with a probability of 0.0032.

New high quality spectroscopy by \citet{cho09} revealed a broad
feature redshifted by 3500-4500 km~s$^{-1}$ in H$\alpha$ and H$\beta$,
which raises a third option that SDSS J1536+0441 is a ``double peaked
emitter'' (DPE), such as seen in Arp\,102B \citep{hal88} and 3C\,390.3
\citep{era94}.  The peculiar emission profile in DPE may originate
from a thin Keplerian gaseous disk (also proposed for SDSS J1536+0441
by Gaskell [2009]). However, as noted by \citet{cho09}, DPE show two
broad components of comparable strength, unlike seen here.

It is clearly important to observe this quasar with sub-arcsecond
resolution, for two reasons.  First, the 0.1-pc binary hypothesis can
be ruled out if two widely-separated sources are detected.  Second, if
observations with sub-arcsecond resolution detect only a single
source, then the random-projection estimate would drop below 0.0032,
serving to strengthen the 0.1-pc binary hypothesis.

SDSS J1536+0441 is not detected in 1.4-GHz sky surveys, so its flux
density is less than 1 mJy at 5\arcsec\, resolution \citep{whi97} and
less than 2.5 mJy at 45\arcsec\, resolution \citep{con98}.  Following
\citet{ive02}, combining the 1 mJy upper limit at 1.4 GHz with the
$i$-band magnitude \citep{bor09} means that SDSS J1536+0441 has a
radio-to-optical ratio less than unity, implying that it is a
radio-quiet quasar.  Fortunately, such a classification does not
exclude it from being detected at radio frequencies and observed with
sub-arcsecond resolution.  \S~\ref{data} of this {\em Letter\/}
reports new VLA\footnote{Operated by the National Radio Astronomy
Observatory, which is a facility of the National Science Foundation,
operated under cooperative agreement by Associated Universities, Inc.}
\citep{tho80} imaging of SDSS J1536+0441 at sub-arcsecond resolution,
leading to the discovery of two sources separated by 0.97\arcsec\,
(5.1 kpc).  The implications of this discovery are explored in
\S~\ref{implications}.  \S~\ref{summary} closes with a summary of this
work and suggestions for future directions.

\section{VLA Imaging}\label{data}

The B configuration of the VLA was used, under proposal code AL738, to
observe SDSS J1536+0441 near transit on UT 2009 February 17 and 20.
Observations were phase-referenced to the calibrator J1534+0131 whose
positional accuracy was less than 2 mas.  The switching angle was
3\arcdeg\, and the switching time was 280~s.  The center frequency was
8.4601~GHz, abbreviated as 8.5~GHz hereafter.  The {\em a priori\/}
pointing position for the quasar was centered 3\arcsec\, north of the
SDSS name-derived position to avoid any phase-center artifacts.
Observations were made assuming a coordinate equinox of 2000.  Data
were acquired with a bandwidth of 100~MHz for each circular
polarization.  Observations of 3C\,286 were used to set the amplitude
scale to an accuracy of about 3\%.  The net exposure times for SDSS
J1536+0441 were 6020~s and 6080~s on 2009 February 17 and 20,
respectively.  Twenty-five antennas provided data of acceptable
quality.

The data were calibrated using the 2009 Dec 31 release of the NRAO
AIPS software.  No polarization calibration or self-calibrations were
performed.  After calibration, each day's visibility data for SDSS
J1536+0441 were concatenated.  The AIPS task {\tt imagr} was applied
to the concatenated data to form and deconvolve images of the Stokes
$I\/$ emission.  A variety of weighting schemes were applied to the
visibility data, converging on the image given in Figure~1 that
optimizes the balance between sensitivity and resolution.

Figure~1 shows two radio sources, labeled VLA-A and VLA-B.
Elliptical-Gaussian fits to those sources in the image plane yielded
the following integrated flux densities, positions, and 1-dimensional
position errors: for VLA-A, $S = 1.17 \pm 0.04$~mJy, $\alpha(J2000) =
15^{h} 36^{m} 36\fs222$, $\delta(J2000) = +04\arcdeg 41\arcmin
27\farcs06$, and $\sigma_{\rm VLA} = 0\farcs1$; for VLA-B, $S = 0.27
\pm 0.02$~mJy, $\alpha(J2000) = 15^{h} 36^{m} 36\fs288$,
$\delta(J2000) = +04\arcdeg 41\arcmin 27\farcs09$, and $\sigma_{\rm
VLA} = 0\farcs1$.  For each source, the flux-density error is the
quadratic sum of the 3\% scale error and the fit residual, while the
position error is the quadratic sum of a term due to the
phase-calibrator position error (less than 0.002\arcsec), the
signal-to-noise ratio (S/N) (less than 0.02\arcsec), and the
phase-referencing strategies (estimated to be 0.1\arcsec).  The image
fits indicated that each source was unresolved and, given the high S/N
data, this was taken to imply a diameter of less than 0.37\arcsec,
corresponding to half of the geometric-mean beam width at FWHM quoted
in Figure~1.  Sources VLA-A and VLA-B have a summed flux density of
$1.44 \pm 0.05$~mJy.  The image fits also yielded a relative source
separation of 0.97\arcsec\, $\pm$ 0.03\arcsec.

\section{Implications}\label{implications}

From the new VLA data, the 8.5-GHz emission from SDSS J1536+0441
consists of two sources, VLA-A and VLA-B, with radio luminosities,
$L_R = \nu L_\nu$, at 8.5 GHz of $5.2 \times 10^{40}$~ergs~s$^{-1}$
and $1.2 \times 10^{40}$~ergs~s$^{-1}$, respectively.  Each source is
unresolved, with a diameter of less than 0.37\arcsec\, (1.9 kpc) and,
as Figure~1 shows, lies within the quasar's optical localization
region \citep{bor09}.  VLA-A and VLA-B are separated by 0.97\arcsec\,
(5.1 kpc).  These could be two related radio sources, both energized
by the candidate 0.1-pc binary system.  Alternatively, VLA-A and VLA-B
could be independent radio sources originating from a binary quasar
system, with a projected separation of 5.1 kpc.  The implications of
these two interpretations are explored separately below in
\S~\ref{related} and \S~\ref{unrelated}.  As mentioned in
\S~\ref{motivation}, SDSS J1536+0441 is a radio-quiet quasar as
defined by \citet{ive02}, and its properties will be analyzed within
that context.  Finally, in \S~\ref{coronal} the available data for
SDSS J1536+0441 will be used to explore a specific framework for
radio-quiet quasars recently proposed by \citet{lao08}.

\subsection{VLA-A and VLA-B Powered by a 0.1-pc Binary}\label{related}

The radio emission from VLA-A and VLA-B could represent two related
sources, with a projected separation of 0.97\arcsec\, (5.1 kpc) and
with both radio sources ultimately being energized by the candidate
0.1-pc binary.  When imaged at sub-arcsecond resolution, other
radio-quiet quasars at similar redshifts and luminosities are known to
exhibit double, triple, and linear radio structures on scales of a few
kiloparsecs \citep{kel94,kul98}, making a radio-double scenario
plausible for J1536+0441.

If either VLA-A or VLA-B coincide with the energizing 0.1-pc binary,
then the localization of that binary could be improved from the
optical diameter of 2\arcsec\, \citep{bor09} to the radio diameter of
less than 0.37\arcsec.  However, by analogy with the \citet{kul98}
radio doubles with accurate optical astrometry, the 0.1-pc binary in
J1536+0441 could be located between VLA-A and VLA-B.  Therefore a
conservative value for the radio localization area is a region
extending about 1\arcsec\, east-west and less than 0.8\arcsec\,
north-south (the major axis of the Gaussian restoring beam), leading
to a radio localization area of less than 0.8 arcsec$^{2}$.  The area
of the optical localization region is 3.1 arcsec$^{2}$ \citep{bor09},
so the new 8.5-GHz imaging has improved the localization by a factor
of at least 3.9.

As mentioned in \S~\ref{motivation}, SDSS J1536+0441 was less than 1
mJy at 1.4 GHz at 5\arcsec\, resolution \citep{whi97}.  In contrast,
the summed flux density for VLA-A and VLA-B is $1.44 \pm 0.05$~mJy at
8.5 GHz (\S~\ref{data}).  Comparing these values suggests that the
overall spectral index for SDSS J1536+0441 is flat, or even rising,
with frequency.  This could indicate that VLA-A, the stronger of the
two 8.5-GHz sources, is sufficiently compact to be synchrotron
self-absorbed and thus resemble components in some other radio-quiet
quasars \citep{bar96,kul98,ulv05}.  Such compact emission could also
hint that VLA-A marks the location of the candidate 0.1-pc binary,
implying an even smaller radio localization for it and also
strengthening its similarities to the radio galaxy 0402+379 that hosts
a 7-pc binary \citep{rod06}.  Some radio-quiet quasars do exhibit time
variability \citep[e.g.,][]{bar05}, so this inference of a flat or
rising overall spectrum for VLA-A is weakened by not having
simultaneous measurements at 1.4 and 8.5 GHz.  But if the spectral
index estimate is inexact due to variability on a decade timescale
then, from causality arguments, the inference about small-scale radio
emission still stands.

\subsection{VLA-A and VLA-B Powered by a 5-kpc Binary}\label{unrelated}

The detection of two point sources at radio frequencies also raises
the possibility that the emission originates from two quasars
0.97\arcsec\, apart, or at a projected separation of 5.1
kpc. \citet{bor09} noted that the probability for such a random
projection in their sample is 0.0032.  Therefore, the two quasars are
most likely not due to a random projection, but are likely physically
related, i.e. a binary quasar system.

A remarkable radio-loud case of a binary system of active galactic
nuclei (AGNs) is seen in 3C\,75, where two systems of two-sided jets
emanate from two close point sources with a projected separation of
$\sim 7.5$~kpc \citep{owe85}.  Compact binary AGNs were also revealed
with {\em Chandra} observations of nearby systems.  Particularly clear
cases are NGC\,6240 with a $\sim 1$~kpc separation \citep{kom03}, and
Mrk\,463 with a $\sim 3.8$~kpc separation \citep{bia08}.  Thus, SDSS
J1536+0441 may be another example of a compact binary AGN system.

A systematic study of the abundance of binary quasars in the SDSS was
carried out by \citet{hen06}, who found a projected correlation
function of the form $(R_{\rm prop}/0.43{\rm Mpc}{\rm
h}^{-1})^{-1.48}$ on scales of 10--40~kpc in proper length (proper
length is used given the absence of a Hubble flow on these small
scales).  Extrapolating to 5~kpc (using $h=0.71$) we get a correlation
function of 1175, i.e. an observed surface density 1175 larger than
expected for random projection.  Using the random-projection estimate
of 0.0032 \citep{bor09}, the expected observed number is actually
3.76, i.e. 3 to 4 such binaries are expected.  The study of Hennawi et
al.  is based on quasars with a mean $z\sim 1.5$ due to the smaller
number of quasars at lower $z$, and it does not extend down to 5~kpc
due to the fiber angular resolution limit.  However, the study of
Hennawi et al. indicates that the probability to find a 5~kpc binary
quasar in the sample used by \citet{bor09} is of the order of unity.

Each of the two quasars in the binary may be contributing its
broad-line emission system to the total light, producing the double
broad-line system discovered by \citet{bor09}.  Thus, rather than
having a binary black-hole system on a scale of 0.1-pc, we may have a
binary quasar system on a 5-kpc scale, most likely residing within two
strongly interacting galaxies.

The velocity separation of 3500 km~s$^{-1}$ \citep{bor09} is rather
large, but not implausible in a cluster of galaxies.  About half the
clusters studied by \citet{car96} show such an extent of velocities.
The maximum velocity differences in the quasar binaries studied by
\citet{hen06} is 1870 km~s$^{-1}$, but that study imposed a cap of
2000 km~s$^{-1}$ on the binary velocity separation.  At a projected
relative velocity of 3500 km~s$^{-1}$, two galaxies cannot form a
bound system, so the term binary quasar here does not refer to a
physically bound binary system.

\subsection{The Radio/X-ray Connection}\label{coronal}

For some small X-ray and radio selected samples of radio-quiet AGNs, a
link between the radio and X-ray emission has been suggested by the
correlation between the radio and the X-ray luminosities
\citep{bri00,sal04,wan06}.  While intriguing, such a radio/X-ray
connection should be verified by using an unbiased survey.  The PG
quasar sample is a complete optically-selected sample \citep{bor92},
making it independent of radio and X-ray biases.  \citet{lao08} have
used these radio-quiet PG quasars to demonstrate that (a) the radio
and the X-ray luminosities are correlated over a large range of AGN
luminosity and (b) the correlation follows $L_R / L_X \sim 10^{-5}$,
the well-established correlation for coronally active cool stars
\citep{gue93}, where $L_R = \nu L_\nu$ at 5 GHz and $L_X$ is in the
0.2-20~keV band.  For cool stars, the $L_R / L_X \sim 10^{-5}$
relation is accepted as a manifestation of coronal heating by
energetic electrons following magnetic reconnection, that subsequently
gives rise to X-ray emission.  By analogy with cool stars,
\citet{lao08} conjecture that the radio emission in radio-quiet AGNs
may also be related to coronal magnetic activity.

This coronal framework can be tested for SDSS J1536+0441 by
making use of its recent Swift observation by \citet{arz09},
carried out in 2009 February 4 and 5, just two weeks before
our VLA observations. \citet{arz09} measure a 0.5-10 keV luminosity of 
$5\times 10^{44}$~erg~s$^{-1}$, with a spectral slope of
$-1.5$. This gives $\nu L_\nu=2.3\times 10^{44}$~erg~s$^{-1}$ at
1 keV, and thus $L_X$ of $1.4\times 10^{45}$~erg~s$^{-1}$ (see \citet{lao08}
for the conversion of $\nu L_\nu$ at 1 keV to $L_X$). The total
radio luminosity we find at 8.5 GHz is $6.4\times 10^{40}$~erg~s$^{-1}$,
which extrapolates to $L_R=8.3\times 10^{40}$~erg~s$^{-1}$, assuming a 
spectral slope of $-0.5$. We therefore get $L_R / L_X = 5.9\times 10^{-5}$
which falls within the range of ratios seen by \citet{lao08} for the
radio-quiet PG quasars. Thus, SDSS J1536+0441 follows the radio/X-ray
relation of optically selected radio quiet quasars.

\section{Summary and Future Directions}\label{summary}

The radio-quiet quasar SDSS J1536+0441 has two broad-line emission
systems that \citet{bor09} interpret as a candidate binary black-hole
system with a separation of 0.1 pc (0.02 mas).  Our new VLA imaging at
8.5 GHz reveals two sources, separated by 0.97\arcsec\, (5.1 kpc),
within the quasar's optical localization region.  Each radio source
has a diameter of less than 0.37\arcsec\, (1.9 kpc).

Other radio-quiet quasars do exhibit double structures, suggesting
that the radio double in SDSS J1536+0441 could be energized by the
candidate 0.1-pc binary.  Alternatively, the radio emission may arise
from a binary system of quasars with a projected separation of 5.1
kpc, and those two quasars may be responsible for the two observed
broad-line emission systems.  Binary AGNs with kpc-scale separations
are known from radio and X-ray observations, and a few such system are
expected in the \citet{bor09} sample based on the observed clustering
of quasars down to a scale of 10 kpc.

Interestingly, the Balmer emission line profiles shown by
\citet{cho09} are noticeably different than those in
\citet{bor09}. For H$\beta$ the the peak of the r component is about
40\% of the peak of b component, as measured above the ``valley''
between the r and b components, while in \citet{bor09} it is about
60\%. Similarly, for H$\alpha$ the r peak is about 75\% of the b peak
in \citet{cho09}, while in \citet{bor09} it is about 95\%. These
could be due to profile variability, but it could also be due to
different contributions of two point sources to the total spectrum,
due to different slit positions in the \citet{bor09} and \citet{cho09}
spectra.

If the 5-kpc binary interpretation is correct then ground-based
adaptive optics imaging in the visible band, with sub-arcsecond
resolution, should clearly show two point sources, and spectroscopy of
those sources should exhibit two different spectra, each one with a single 
broad emission-line system.  Deep optical imaging on arcsecond scales may
also reveal the nature of the host galaxy, and whether strongly
interacting galaxies are involved. 

Deep optical imaging on arcminute scales can establish whether or not
SDSS J1536+0441 resides in a cluster.  \citet{bor09} noted two nearby
galaxies with similar photometric redshifts, which may indicate the
presence of such a cluster.  However, a cluster is likely in both the
0.1-pc and 5-kpc binary interpretations, as a binary black-hole also
requires an earlier merger.  The only difference between the two
interpretation is the timescale since the merger occurred, which can be
a fraction of the Hubble time for a 0.1-pc binary.  On such a
timescale the immediate environment may relax, but the larger scale
cluster environment evolves independently of the merger process.

Follow-up radio observations at other frequencies and epochs, as well
as with higher angular resolutions, can help distinguish between the
0.1-pc and 5-kpc binary interpretations.  Specifically, demonstrating
that each radio source has a flat spectrum, is time-variable, and
remains compact at higher resolution would strongly support the 5-kpc
binary interpretation.  In contrast, in the 0.1-pc binary case, one
(or both) of the radio sources would be expected to be steady in time,
have a steep spectrum, and begin to show an outflow-like structure
when examined with higher resolution; a flat-spectrum source paired
with a steep-spectrum source would suggest a core-jet morphology.

If the 5-kpc binary interpretation is correct, then SDSS J1536+0441
may form the first known compact binary quasar of two luminous
broad-line systems, unlike the earlier X-ray-discovered compact
systems where one or both of the AGNs are partly obscured, or broad
emission lines are missing altogether.

\acknowledgments
We thank Shai Kaspi for his help with the X-ray analysis.  We
acknowledge useful feedback from two anonymous referees.  This
research was supported by THE ISRAEL SCIENCE FOUNDATION (grant
\#407/08), and by a grant from the Norman and Helen Asher Space
Research Institute.

{\it Facilities:} \facility{VLA}

\clearpage

\begin{figure}
\epsscale{1.0}
\plotone{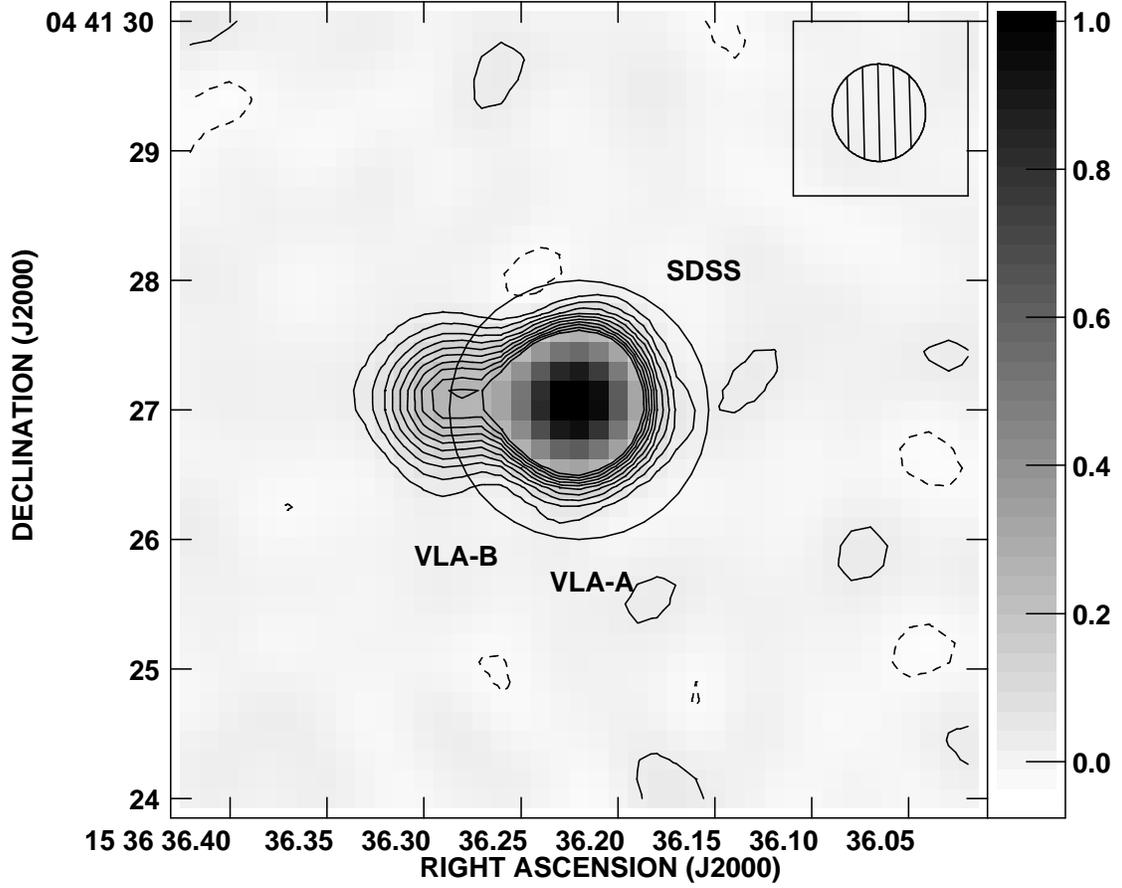}
\caption{VLA image of Stokes $I\/$ emission from SDSS J1536+0441 at a
frequency of 8.5~GHz and spanning 6\arcsec\, (31~kpc).  SDSS position
is marked with a circle of radius 1\arcsec.  Labels identify sources
VLA-A and VLA-B.  Uniform weighting with a robustness parameter of
zero was used, giving an rms noise of 0.013~mJy~beam$^{-1}$ (1
$\sigma$) and Gaussian beam dimensions at FWHM of 0.75\arcsec\, times
0.72\arcsec\, at a position angle of 1.3\arcdeg\, (hatched ellipse).
Geometric-mean beamwidth is 0.73\arcsec\, (3.8 kpc) at FWHM.  Contours
are at $-$6, $-$4, $-$2, 2, 4, 6, 8, 10, 12, ...  20 times 1 $\sigma$.
Negative contours are dashed and positive ones are solid.  Image peak
is 1.15~mJy~beam$^{-1}$.  Linear gray scale spans
$-$0.05~mJy~beam$^{-1}$ to 1.0~mJy~beam$^{-1}$.}\label{fig1}
\end{figure}
\clearpage


\begin{thebibliography}{}
\bibitem[Arzoumananian et al.(2009)]{arz09} Arzoumananian, Z.,
 Lowenstein, M., Mushotzky, R. F., \& Gendreau, K. C. 2009, ATel 1931
\bibitem[Barvainis et al.(1996)]{bar96} Barvainis, R., Lonsdale, C.,
 \& Antonucci, R. 1996, \aj, 111, 1431
\bibitem[Barvainis et al.(2005)]{bar05} Barvainis, R., Lehar, J.,
 Birkinshaw, M., Falcke, H., \& Blundell, K. M. 2005, \apj, 618, 108
\bibitem[Bianchi et al.(2008)]{bia08} Bianchi, S., Chiaberge, 
M., Piconcelli, E., Guainazzi, M., \& Matt, G.\ 2008, \mnras, 386, 105 
\bibitem[Boroson \& Green(1992)]{bor92} Boroson, T.A., \& Green,
 R. F. 1992, \apjs, 89, 109
\bibitem[Boroson \& Lauer(2009)]{bor09} Boroson, T.A., \& Lauer, T. R.
 2009, \nat, in press (astro-ph/0901.3779)
\bibitem[Brinkmann et al.(2000)]{bri00} Brinkman, W., et al. 2000,
 \aap, 356, 445
\bibitem[Carlberg et al.(1996)]{car96} Carlberg, R.~G., Yee, 
H.~K.~C., Ellingson, E., Abraham, R., Gravel, P., Morris, S., 
\& Pritchet, C.~J.\ 1996, \apj, 462, 32 
\bibitem[Chornock et al.(2009)]{cho09} Chornock, R., et al. 2009,
 ATel 1955
\bibitem[Condon et al.(1998)]{con98} Condon, J. J., Cotton, W. D.,
 Greisen, E. W., Yin, Q. F., Perley, R. A., Taylor, G. B., \&
 Broderick, J. J. 1998, \aj, 115, 1693
\bibitem[Eracleous \& Halpern(1994)]{era94} Eracleous, M., \& Halpern,
 J. P. 1994, \apjs, 90, 1
\bibitem[Gaskell(2009)]{gas09} Gaskell, M. 2009, \nat, submitted
\bibitem[Guedel \& Benz(1993)]{gue93} Guedel, M., \& Benz, A.O. 1993,
 \apj, 405, L63
\bibitem[Halpern \& Filippenko(1988)]{hal88} Halpern, J. P., \&
 Filippenko, A. V. 1988, \nat, 331, 46
\bibitem[Hennawi et al.(2006)]{hen06} Hennawi, J.~F., et al. 2006, \aj,
 131, 1
\bibitem[Ivesic et al.(2002)]{ive02} Ivesic, Z., et al. 2002, \aj,
 124, 2364
\bibitem[Kellermann et al.(1994)]{kel94} Kellermann, K. I., Sramek,
 R. A., Schmidt, M., Green, R. F., \& Shaffer, D. B. 1994, \aj, 108,
 1163
\bibitem[Komossa et al.(2003)]{kom03} Komossa, S., Burwitz, 
V., Hasinger, G., Predehl, P., Kaastra, J.~S., 
\& Ikebe, Y.\ 2003, \apjl, 582, L15 
\bibitem[Kukula et al.(1998)]{kul98} Kukula, M. J., Dunlop, J. S.,
 Hughes, D. H., \& Rawlings, S. 1998, \mnras, 297, 366
\bibitem[Laor \& Behar(2008)]{lao08} Laor, A., \& Behar, E. 2008,
 \mnras, 390, 847
\bibitem[Owen et al.(1985)]{owe85} Owen, F.~N., O'Dea, C.~P., Inoue,
M., \& Eilek, J.~A. 1985, \apj, 294, L85
\bibitem[Rodriguez et al.(2006)]{rod06} Rodriguez, C., Taylor, G. B.,
 Zavala, R. T., Peck, A. B., Pollack, L. K., \& Romani, R. W. 2006,
 \apj, 646, 49
\bibitem[Salvato et al.(2004)]{sal04} Salvato, M., Greiner, J., \&
 Kuhlbrodt, B. 2004, \apj, 600, L31
\bibitem[Thompson et al.(1980)]{tho80} Thompson, A. R., Clark, B. G.,
 Wade, C. M., Napier, P. J. 1980, \apjs, 44, 151
\bibitem[Ulvestad et al.(2005)]{ulv05} Ulvestad, J. S., Antonucci,
 R. R., \& Barvainis, R. 2005, \apj, 621, 123
\bibitem[Wang et al.(2006)]{wan06} Wang, R., Wu, X.-B., \& Kong,
 M.-Z. 2006, \apj, 645, 890
\bibitem[White et al.(1997)]{whi97} White, R. L., Becker, R. H.,
 Helfand, D. J., \& Gregg, M. D. 1997, \apj, 475, 479
 (http://sundog.stsci.edu/top.html)
\end{thebibliography}
\end{document}